\documentclass[aps,preprint,showpacs]{revtex4}
\usepackage{bm}
\usepackage{graphics}
\usepackage{graphicx}
\begin{document}
\title{Effect of genome sequence on the force-induced unzipping of a DNA molecule}
\author{Navin Singh and Yashwant Singh }
\affiliation{Department of Physics, Banaras Hindu University, Varanasi - 221 005, India}
\begin{abstract}
We considered a dsDNA polymer in which distribution of bases are random at the base pair level
but ordered at a length of 18 base pairs and calculated its force elongation behaviour in the
constant extension ensemble. The unzipping force $F(y)$ vs. extension $y$ is found to have a
series of maxima and minima. By changing base pairs at selected places in the molecule we calculated
the change in $F(y)$ curve and found that the change in the value of force is of the order of
few pN and the range of the effect depending on the temperature, can spread over several base pairs. 
We have also discussed briefly how to calculate in the constant force ensemble a pause or a jump in 
the extension-time curve from the knowledge of $F(y)$.
\end{abstract}
\maketitle

\section{Introduction}

DNA is a giant double stranded linear polymer in which genetic information is stored \cite{watson}. 
Its double stranded helical structure is stabilized by the hydrogen bonding between complimentary 
bases (A-T are linked with two hydrogen bonds and G-C by three hydrogen bonds) and the stacking 
interactions of the base pair plateaux. The stacking interactions which impose a well defined 
distance between the bases and give rise high rigidity to the polymer along its axis depend on the 
genome sequence \cite{saenger,calldine}. The energy landscape of a dsDNA polymer is therefore 
expected to depend on the arrangement of bases along the two strands. Knowing this dependence is an 
important step in understanding the biological functioning of DNA.

With the development of single molecule techniques \cite{smith,amb,florin,cluzel,strick} 
it has now become possible to probe the force elongation characteristics of a double stranded 
DNA (dsDNA) polymer and measure its response to an external force or torque {\it in vitro} at
temperatures where dsDNA is thermally stable. Such measurements give informations about the 
energy landscape of the molecule. Experiments have usually been performed either in the constant 
extension \cite{bock} or in the constant force ensemble \cite{danil}. In the constant extension 
ensemble the average force of unzipping is found to vary randomly about an average value as the 
extension is increased \cite{bock}, while in the constant force ensemble the unzipped length as a 
function of time is found to show several pauses and long jumps \cite{danil}.

A number of theoretical efforts have recently been made to understand various aspects of dsDNA
unzipping \cite{danil,thomp,somen,cocco,ns3,nelson}. It is shown that while a homogeneous 
dsDNA gains considerable entropy by opening in response to the external
force and therefore the unzipping is entropy driven, a heterogeneous dsDNA is believed to
unzip primarily for energetic reasons \cite{nelson}.
Lubensky and Nelson \cite{nelson} have studied the force induced unzipping of a randomly
disordered dsDNA using a Hamiltonian which is coarse grained over many but unknown number of
bases. Weeks et al \cite{danil} have used the model of 
Lubensky and Nelson \cite{nelson} and have calculated the pause point spectrum in the
constant force ensemble of a $\lambda$ phage DNA. 

Our aim in this article is to use a Hamiltonian which describes interactions
at the base pair level and show that the force-extension curve obtained in the constant
extension ensemble provides a more direct exploration of the underlying free energy
landscape from the maxima and minima of the force profile. The model described in Sec. 2
also allows us to calculate the effect of base pair mutation on the force extension 
behaviour.

We consider a dsDNA polymer of $N(=Mn)$
base pairs made by repeating $M(M\to \infty)$ times an oligonucleotide of $n$ base
pairs. The oligonucleotide used here to construct the dsDNA polymer has 18 base pairs of which
9 are A-T(or T-A) and other 9 are G-C(or C-G). The arrangement of these base pairs in the 
oligonucleotide is as shown below;
\begin{eqnarray*}
& {\rm 3'-AGTGACATACTCGGACGA-5'} \\
& {\rm 5'-TCACTGTATGAGCCTGCT-3'}
\end{eqnarray*}
The dsDNA polymer constructed in this way is heterogeneous as it contains both A-T and G-C base 
pairs. However, because of the repetition of the oligonucleotide the distributions of bases in the 
dsDNA polymer are not random but have a periodicity at the length of the oligonucleotide. 
We therefore expect its properties to lie in between a homogeneous \cite{cocco,ns3} and a randomly 
disordered \cite{nelson} dsDNA polymer.

\section{The Model and its Thermodynamics}

We represent the interactions in the polymer at the base pair level by the model Hamiltonian of 
Peyrard and Bishop (PB) \cite{pb} which though ignores the helicoidal structure of the dsDNA 
polymer, has enough details to analyze mechanical behaviour at few \AA\ scale relevant 
to molecular-biological events \cite{pbrev} and can easily be extended to include the
effect of heterogeneity in the base pair sequence. 
The PB model for a heterogeneous DNA polymer is written as,
\begin{equation}
H = \sum_i \left[ \frac{p_i^2}{2m} + V_i(y_i) + W(y_i,y_{i+1})\right]
\end{equation}
where $m$ is the reduced mass of a base pair (taken to be same for both A-T and G-C base pairs
and equal to 300 a.m.u.), $y_i$ denotes the stretching of the hydrogen bonds connecting the two 
bases of the $i^{th}$ pair and $p_i = m(dy_i/dt)$. The on-site potential $V(y_i)$ which describes 
interactions of two bases of the $i^{th}$ pair is represented by the Morse potential
\begin{equation}
V_i(y_i) = D_i[e^{-a_iy_i}-1]^2
\end{equation}
where $D_i$ measures the depth of the potential and $a_i$ its range. Both $D_i$ and $a_i$ depend on 
whether the $i^{th}$ base pair is A-T or G-C. The stacking interaction term of the PB
model \cite{pb} is modified and is written as
\begin{equation}
W(y_i,y_{i+1}) = \frac{1}{2} k \left[1 + \rho_{i,i+1}e^{-b(y_i + y_{i+1})}\right](y_i - y_{i+1})^2
\end{equation}
where force parameter $k$ measures the stiffness of a single strand of the molecule and the second 
term in the bracket represents the anharmonic term. The strength of anharmonic term is measured by 
$\rho_{i,i+1}$ and its range by $b$. We allow the value of the parameter 
$\rho_{i,i+1}(={\bar \rho} + \Delta \rho_{i,i+1})$ to depend on the arrangement of four bases in the 
consecutive base pairs $i$ and $(i+1)$. In our calculation we have taken $D_{AT} = 0.058 \; {\rm eV},
D_{GC} = 0.087 \; {\rm eV}, a_{AT} = 4.2 \; {\rm \AA^{-1}},a_{GC} = 6.3 \; {\rm \AA^{-1}},
b = 0.35 \; {\rm \AA^{-1}}, k = 0.02 \; {\rm eV \AA^{-2}}, {\bar \rho} = 5.0 $ and 
used the data of stacking energies given by Sponer et al \cite{hobza} to estimate the
value of $\Delta \rho_{i,i+1}$. We list in Table 1 the value of $\Delta \rho_{i,i+1}$ for all
possible combinations of two consecutive base pairs.
\begin{table}
\begin{center}
\caption{Value of $\Delta \rho_{i,i+1}$ for all possible combination of two consecutive base
pairs. Two consecutive bases along a strand is shown. Other two bases of the quartet 
are complementary to these. These values are found using the data of stacking energy of 
Sponer et al \cite{hobza} and taking ${\bar \rho} = 5.0 $.}
\vspace{0.5cm}
\begin{tabular}{|l|r|l|r|} \hline\hline
{\bf Base Quartet} & $\Delta \rho_{i,i+1}$ & {\bf Base Quartet} & $\Delta \rho_{i,i+1}$ \\
(3'-5') & & (3'-5') &  \\ \hline
AA & -0.10 & GA & -0.10 \\ \hline
AT & -0.77 & GT & -0.14 \\ \hline
AG & -0.53 & GG & -0.34 \\ \hline
AC & -0.14 & GC &+2.39 \\  \hline
TA & -0.53 & CA & +0.36 \\ \hline
TT & -0.10 & CT & -0.53 \\ \hline
TG & +0.36 & CG & +0.48 \\ \hline
TC & -0.10 & CC & -0.34  \\ \hline \hline
\end{tabular}
\end{center}
\end{table}
We treat the nucleotide the repetition of which forms the dsDNA polymers as an effective
base pair and define its kernel as,
\begin{eqnarray}
{\bar K(y_1,y_{n}}) &=& \int dy_2,....dy_{n-1}\; K(y_1,y_2),..., \nonumber \\
&& \times K(y_i,y_{i+1}),...,K(y_{n-1},y_n) 
\end{eqnarray}
and
$$
H(y_i, y_{i+1}) = \frac{1}{2}\left[V(y_i) + V(y_{i+1})\right] + W(y_i, y_{i+1})
$$
where \cite{ns3}
\begin{equation}
K(y_i,y_{i+1}) = \left(\frac{\beta k}{2\pi}\right)^{1/2} \exp[-\beta H(y_i,y_{i+1})]
\end{equation}
Equation (4) is evaluated by the method of matrix multiplication. For this we chose
$-5.0 \; {\rm \AA}\; {\rm and}\; 200.0 \; {\rm \AA}$, respectively, as the lower
and upper limits of integration for each variable and discretized the space using the
Gaussian quadrature with number of grid points equal to 900. The resulting matrix 
${\bar K(y_1,y_n)}$ is a $900 \times 900 $ square matrix.

The configurational partition function $Z^c_N$ defined as \cite{ns3},
\begin{equation}
Z_N^c = \int \prod_{p=1}^M dy_p {\bar K(y_p,y_{p+1})} \delta(y_1 - y_{N+1})
\end{equation}
has been evaluated by the matrix multiplication method for several values of $N$ ranging
between 3000 to 6000. As all matrices in Eq.(6) are identical the multiplication is
done very efficiently. The resulting partition function is used to calculate the free
energy per base pair from the following relation 
\begin{equation}
f = -\frac{1}{2}k_B T\ln\left(\frac{4\pi^2k_B^2T^2 m}{k}\right) - \frac{k_B T}{N}\ln Z_N^c
\end{equation}
where the first term on the r.h.s. is due to the kinetic energy. We found that for 
$N \geq 4000 $ the value of $f$ is independent of the value of $N$ taken in the calculation
of $Z_N^c$.  

We have also diagonalized the matrix ${\bar K(y_p,y_{p+1})}$ to find first two eigenvalues 
$\lambda_0 $ and $\lambda_1 $ where $\lambda_0 = \exp(-\beta E_0)$ and $\lambda_1 = \exp(-\beta E_1)$,
using a method described in ref \cite{nrc}. Since $E_0$ and $E_1$ are the eigenenergies of a 
kernel having $n=18$ base pairs, the average eigenenergies per base pair is $\epsilon_i = E_i/n$.
In Fig. 1(a) we plot the eigenenergies $\epsilon_0$ and $\epsilon_1$. We find that 
$\Delta \epsilon(T) = \epsilon_0 - \epsilon_1 \propto (T_D - T)^{\nu}$ with $\nu = 1$ and
$T_D = 356.7$ K. The behaviour of $\Delta \epsilon(T)$ as a function of $T$ is found to be same
as in the case of a homogeneous dsDNA polymer \cite{ns3}. In the thermodynamic limit the value of
configurational partition function $Z_N^c$ is determined by $\lambda_0$ and therefore 
$Z_N^c = \lambda_0^M = \exp(-\beta N \epsilon_0)$. The free energy per base pair calculated using 
this value of $Z_N^c$ agrees very well with the values found from Eq.(7).

The value of $f$ as a function of temperature $T$ is shown in Fig. 1(b). A cusp in $f$ at the thermal 
denaturation temperature $T_D = 356.7$ K is clearly seen. The existence of cusp indicates that 
the thermal denaturation transition is first order with a sudden jump in its entropy.
\begin{figure}[h]
\begin{center}
\includegraphics[width=3.0in,height=4.5in]{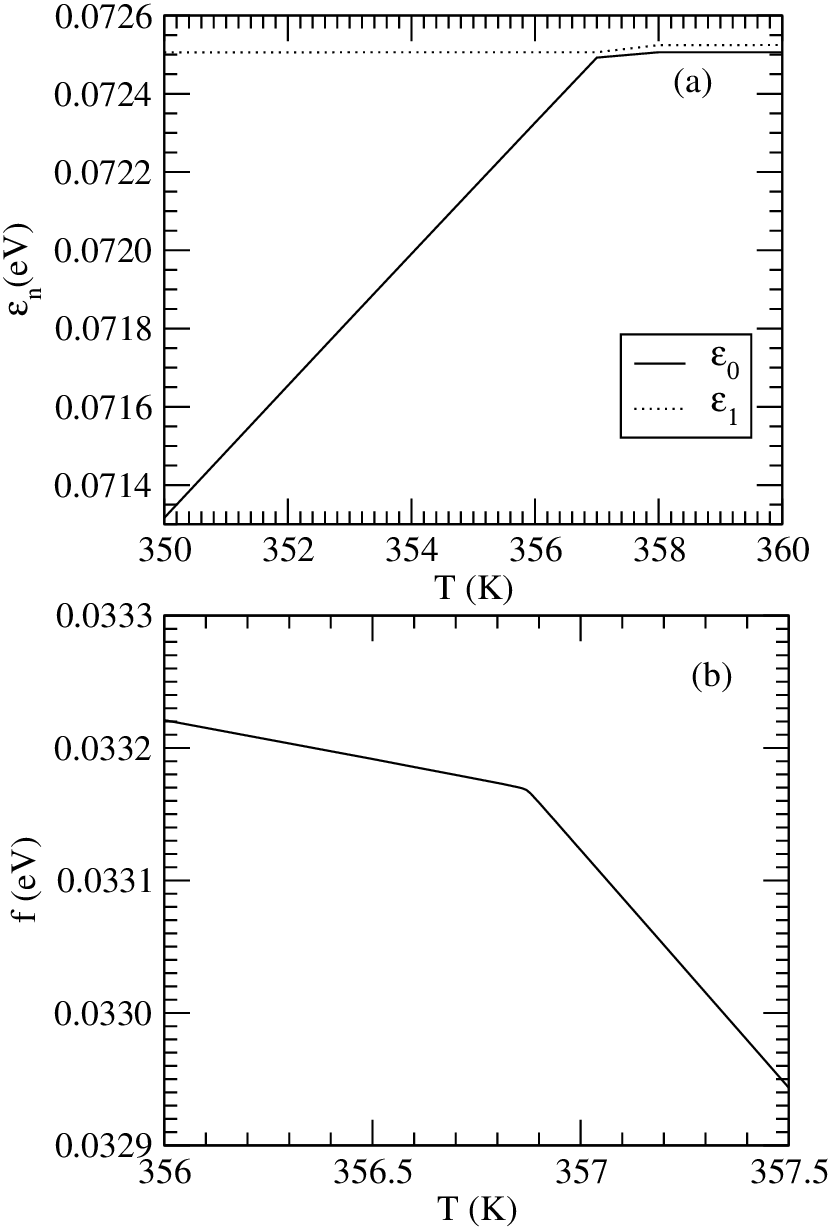}
\caption{\small (a) The average per base pair eigenenergies $\epsilon_0$ and $\epsilon_1$ of the 
kernel of Eq.(4) as a function of temperature are shown. (b) The free energy per base pair as a 
function of temperature is shown. A cusp in the free energy is seen at the thermal denaturation 
temperature $T=T_D =356.7$ K where $\Delta \epsilon = \epsilon_0 - \epsilon_1 \sim 0$.}
\end{center}
\end{figure}

Next we calculate the unzipping of the polymer in the constant extension ensemble in 
which separation of one ends of the two strands of the molecule is kept fixed and
the average force needed to keep this separation is measured. The work done in stretching
the base pair $1$ to $y$ distance apart is \cite{ns3} 
\begin{equation}
W(y) = \frac{1}{2} V_1(y) - k_B T [\ln Z_n^c(y) - \ln Z_N^c]
\end{equation}
where
\begin{equation}
Z_N^c(y) = \int \prod_{p=1}^M dy_p \delta (y_1 - y) \delta (y_{N} - 0){\bar K}(y_p,y_{p+1})
\end{equation}
and $Z_N^c$ is defined by Eq.(6). The force $F(y)$ as a function of extension $y$
is found from the relation,
\begin{equation}
F(y) = \frac{\partial W(y)}{\partial y}
\end{equation}

\section{Results and Discussions}

In Fig. 2 we plot the value of $F(y)$ as a function of extension $y$ for $T = 100$ K and 300 K. 
To show the height and width of peak in $F(y)$ at small values of $y$ as well as oscillations at 
larger values of $y$ we chose different scales on the two sides of $y=5.0 \; {\rm \AA}$. Though the 
experiments are generally done at temperatures close to 300 K, the motivation for studying the 
behaviour of dsDNA at 100 K is to illustrate the effect of temperature on the energy landscape. 
Figure 2 shows a large force barrier at $y \sim 0.2 \; {\rm \AA}$, a feature 
similar to that found in the case 
of a homogeneous dsDNA polymer \cite{cocco,ns3}. The height of this peak is nearly 230 pN at 100 K 
and 215 pN at 300 K. The physical origin of this barrier is in the potential well
due to hydrogen bonding plus the additional barrier associated with the reduction in DNA
strand rigidity as one passes from dsDNA to ssDNA. Since the peak corresponds to a process
in which only one or two base pairs participate the effect of thermal energy in formation of the 
barrier is small.

\begin{figure}
\begin{center}
\includegraphics[width=4.5in,height=3.0in]{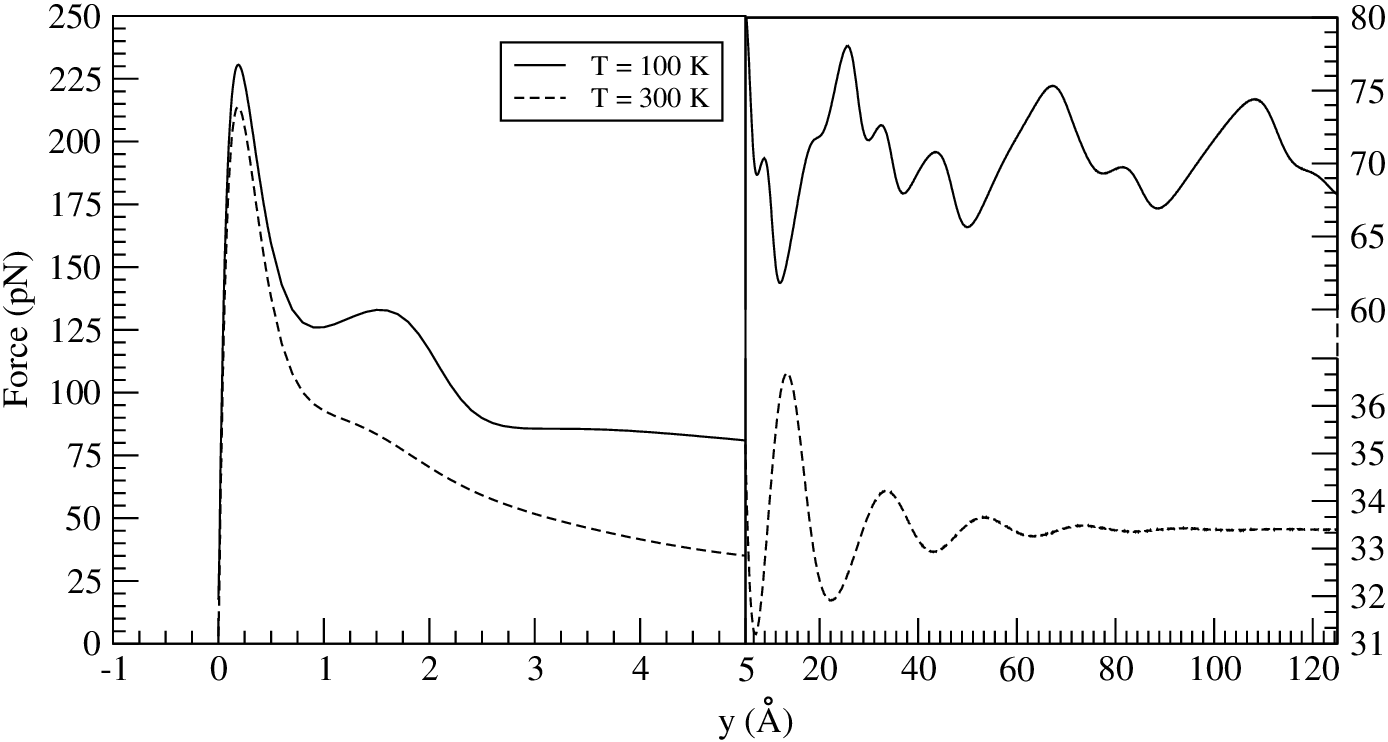}
\caption{\small The average force $F(y)$ in pN at $T = 100 $ K and 300 K required to stretch
one of the ends base pair to a distance $y$ is shown. Different scales are chosen for the two
sides of $y=5.0 \; {\rm \AA}$.}
\end{center}
\end{figure}

For $y \geq 1.0 \; {\rm \AA}$ we, however, find that the two curves of $F(y)$ 
considerably differ from each other as well as from that of a homogeneous DNA polymer.
While in case of a homopolymer the peak in $F(y)$ decays as $y$ is increased and for
$y \geq 10.0 \; {\rm \AA}$ attains a constant value equal to that of the critical
force found from the constant force ensemble to unzip the dsDNA into two ssDNA
\cite{ns3}, here we find that $F(y)$ curve oscillates about a mean value.
These oscillations are due to maxima and minima in the energy landscape and these
maxima and minima depend on the genome sequence in the two strands of the DNA polymer.
It is easy to realize that a G-C rich region of the polymer has energy minimum whereas
the A-T rich region has energy maximum. Therefore, when the front of the unzipping
fork enters the G-C rich region it needs larger force to come out of it whereas
in case of the A-T rich region it needs less force than the average to move forward.
As is evident from Fig. 2, the maxima and minima in the energy landscape are much
more pronounced at 100 K compared to that at 300 K; the thermal fluctuations have
tendency to suppress the local variation and make the energy landscape relatively smooth.

At $T$ = 300 K the oscillations in $F(y)$ are found to decay (see Fig. 2) rather
rapidly and for extension greater than 100 ${\rm \AA}$ the dsDNA polymer seems
to behave like a homopolymer in which unzipping takes place continuously at constant
rate when the applied force exceeds the critical force. But at $T$ = 100 K the
oscillations in $F(y)$ persists for much larger values of $y$. As the extension
$y$ increases, however, the wiggles in $F(y)$ get smoothened and the peak heights
decrease though very slowly. These features arise due to the contribution made to the
free energy by the fluctuations of the single stranded part of the unzipping fork.
As $y$ increases the single stranded part increases resulting in larger entropic
contributions to the free energy and thus reducing the barrier encountered by the front
of the fork. After certain length of the ssDNA part the entropic contribution to free
energy per base pair gets saturated and oscillations in $F(y)$ if not already suppressed
will remain unaffected on further increasing the extension. Therefore the effect of
genome sequence on the unzipping depends on the depth of the local energy minimum.
In the case of the dsDNA polymer considered here the variations in the energy along the polymer
gets averaged out at 300 K and therefore the unzipping beyond about 100 base pairs becomes
similar to that of a homopolymer. But at 100 K the local minimum in the energy landscape are deep 
enough to show variations in the unzipping force for large enough extensions.
\begin{figure}[h]
\begin{center}
\includegraphics[width=3.0in,height=3.5in]{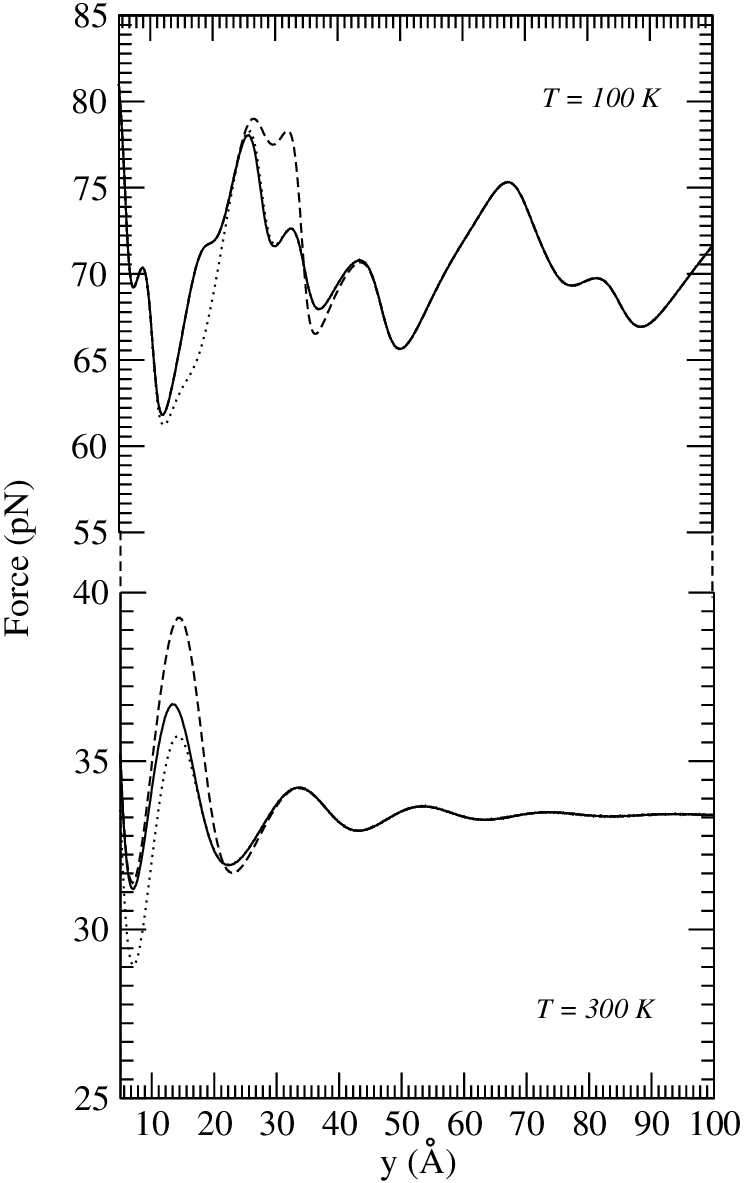}
\caption{\small The change in $F(y)$ when a base pair C-G at position 10 is changed by a base pair
T-A (dotted line) and when a base pair A-T at position 15 is changed by a base pair G-C (dashed line)
is shown at temperatures 100 K and 300 K. The full line corresponds to the original dsDNA polymer.
This change is due to the combined effect of the change in the on-site and the stacking potentials.}
\end{center}
\end{figure}

To examine more closely the sensitivity of the force - extension curve on the genome
sequence we altered base pairs on selected positions and calculated their effects
on the $F(y)$ curve. We indicate a base pair by its number counted from the stretched
end taking the base pair that is being stretched as 1. 
First we alter a base pair at one position and calculate its effect. In Fig. 3
we show our results for two cases, (i) a base pair C-G at position 10 is replaced
by a base pair T-A (shown by dotted line) and (ii) a base pair A-T at position 15
is replaced by a base pair G-C (shown by dashed line). These replacements have brought
changes in both the on-site potential $V(y)$ and in the stacking interactions. The
change in the stacking interactions is measured by the change in $\Delta \rho$ which
for base pairs 9-10 has changed from -0.14 to -0.77 and for the base pairs 10-11 from
-0.53 to -0.10. When the base pair at position 15 has been changed from A-T to G-C
the value of $\Delta \rho$ has changed for base pairs 14-15 from -0.10 to -0.34 and for
base pair 15-16 from -0.14 to +2.39. The change in the $F(y)$ curve brought by the change in
the base pair sequence is therefore due to the combined effect of the change in the on-site and the
stacking potentials. From Fig. 3 we note that the change in $F(y)$ value is of the
order of 5-7 pN and the range of the effect spreads about 14-17 ${\rm \AA}$. This means
that while the effect is localized to about 7-8 base pairs at 100 K, at 300 K it spreads
to almost the length of the oligonucleotide.
\begin{figure}
\begin{center}
\includegraphics[width=3.0in,height=3.5in]{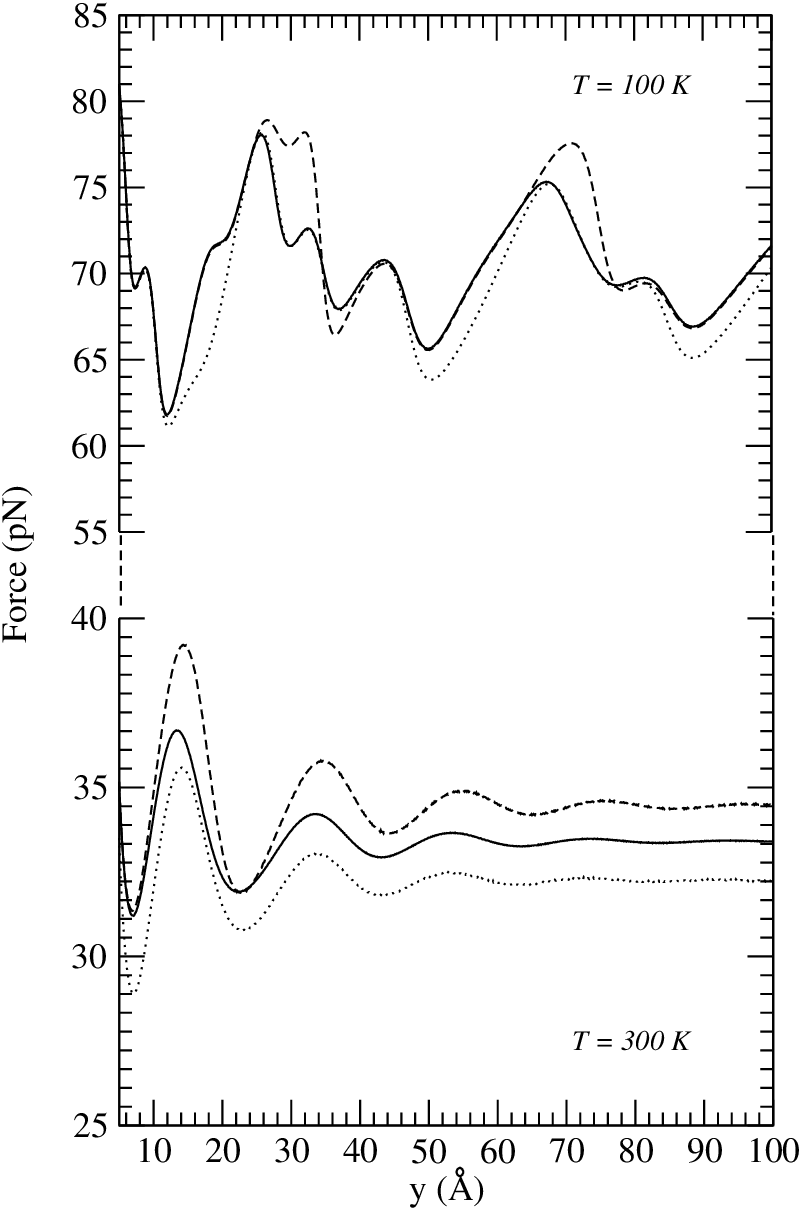}
\caption{\small The change in the value of $F(y)$ when a base pair in each repeating nucleotide is
introduced at $T = 100 $ K and 300 K. The dotted line corresponds to the change introduced at
positions 10, 28, 46, \ldots by replacing C-G base pair by T-A base pair. Similarly the dashed line
corresponds to the situation when the base pair A-T from positions 15, 33, 51, \ldots is replaced
by the G-C base pair. The full line corresponds to the original dsDNA polymer.}
\end{center}
\end{figure}

In Fig. 4 we compare the results found by replacing a base pair in each repeating 
nucleotide at the same position, i.e. a periodic change in the base pair sequence
with the length of periodicity equal to that of the oligonucleotide. For example,
the change introduced at positions 10 now repeats along the polymer at positions 
10, 28, 46, 64,... and the change made at position
15 now repeats along the polymer at positions 15, 33, 51, 69,.. While the change introduced
in the $F(y)$ curve due to this change in genome sequence follow the periodicity
of the change at 100 K, at 300 K the entire curve either moves up or down by about
1 pN for $y \geq 50 \; {\rm \AA}$.

\section{Conclusions}

We have modified the stacking energy part of the PB Hamiltonian so that the effect arising
due to genome sequence in a dsDNA molecule is fully accounted for. Using the method of
matrix multiplications we have done essentially exact thermodynamics of this Hamiltonian
in the constant extension ensemble.
The results given above suggest that the genome sequence has a very specific effect on the unzipping
of a dsDNA polymer and can therefore be used to find this sequence by determining the force
extension curve in the constant extension ensemble. 
Any change in this sequence can change the force-extension curve along 
the length of few base pairs. Such a change in unzipping process may have an important effect
on the DNA transcription and replication dynamics.

The above results can also be used to calculate the unzipping properties
of the molecule in the constant force ensemble. For example, the constant force ensemble partition
function can be found from the relation,
\begin{equation}
Z_N^c(F) = \int dy \; Z_N^c(y) \; e^{\beta F y} = Z_N^c \int dy \; e^{\beta(Fy - W(y))}
\end{equation}
and the time needed to cross a force peak (or valley) encountered by the front of the unzipping 
fork from the relation
\begin{equation}
t=t_0 \; \exp\left(\beta \int_{y_1}^{y_2} dy' (F(y') - F)\right)
\end{equation}
where $F(y_1) - F = F(y_2) - F = 0$ and $t_0$ is time needed by the front to move the distance;
$\Delta y = y_2 - y_1$ under the influence of the force $F$ when there is no peak or valley. 
A natural extension of the method developed here is to apply it to estimate the 
force-elongation behaviour of a natural DNA \cite{ns4}.

Experimental results both in the constant extension and in the constant force ensemble are
available for lambda phage DNA which is 48,502 base pairs long. This DNA is particularly
interesting as it consists of a GC rich half connected to an AT rich half and therefore
expected to have different energy landscape viewed from the two ends. Though for
quantitative comparisions between the experimental results and the theory one has
to consider the genome sequence of the DNA from the end used in the experiment, the 
qualitative features of the experimental results are in agreement with the features
of the force-extension curve discussed above. It may also be pointed out that
while the method described above give the force behaviour at the base pair level the
results found experimentally are still at a level of several base pairs.

\section{Acknowledgments}

The work was supported by research grant from the Department of Science \& Technology (DST), India.
One of us (N.S.) is thankful to Council of Scientific \& Industrial Research (CSIR), India for
research fellowship. We are thankful to Dr. S. Kumar for helpful discussions.


\end{document}